\documentstyle[epsfig]{qcdparis}
\begin{document}
\pagestyle{plain}
\title{QCD Analysis of Diffractive DIS at HERA}

\author{K. Golec--Biernat$^\ast$  and J. Kwieci\'nski}

\affil{H. Niewodnicza\'nski Institute of Nuclear Physics \\
       ul.Radzikowskiego 152, 31-342 Krak\'ow, Poland}

\abstract{
The QCD analysis of deep inelastic diffractive scattering at HERA is performed
assuming the dominance of the "soft" pomeron exchange and simple, physically
motivated parametrization of parton distributions in the pomeron.
 Both the LO and NLO approximations are considered and the
theoretical predictions concerning the quantity $R=F_L^D/F_T^D$
for diffractive structure functions are presented.
}

\resume{
Une analyse de la diffusion diffractive profond\'ement in\'elastique \`a
HERA est present\'ee, bas\'ee sur la dominance d\'{}un pom\'eron "doux" et une
simple,
physiquement motiv\'ee param\'etrisation des distributions des partons dans
le pom\'eron. Les approximations des logarithmes dominants  et au-del\'a
des logarithmes  dominants  ont et\'ees consider\'ees et
les predictions th\'eoriques pour
la quantit\'e $R=F_L^D/F_T^D$ pour les fonctions diffractives de structure
sont
present\'es.
}
\twocolumn[\maketitle]
\fnm{7}{Talk given in the
diffractive  interaction  session
at the Workshop on Deep Inelastic Scattering and QCD,
Paris, April 1995}

%%%%%%%%%%%%%%%%%%%%%%%%%%%%%%%%%%%%%%%%%%%%%%%%%%%%%%%%%%%%%%%%
The main aim of this paper is to analyse the diffractive processes in deep
inelastic $ep$ scattering at HERA \cite{ZEUS,H11,H12},
assuming that they are dominated by
a "soft" pomeron exchange  with the pomeron being described as
a Regge pole with its trajectory
\begin{equation}
\alpha_P(t)=\alpha_P(0) + \alpha_P^{\prime}~t  ~,
\label{trajpom}
\end{equation}
where   $\alpha_P(0) \approx 1.08$ and $\alpha_P^{\prime}
=0.25$ GeV$^{-2} $\cite{DOLA1}-\cite{COLLINS} .
Different  description of these processes  based on
the  "hard" pomeron which follows from perturbative QCD
has been discussed in  \cite{KOLYA}-\cite{BARTELS2}.

The diffractive structure functions have the following factorizable expression
:
\begin{equation}
{dF_{2,L}^D(x_P,\beta,Q^2,t)\over dx_P dt}=f(x_P,t)~F^P_{2,L}(\beta,Q^2,t) ~.
\label{factor1}
\end{equation}
 The function $f(x_P,t)$ is the so called "pomeron flux factor" which,
 if the diffractively recoiled system is a single proton,
has the following form \cite{COLLINS}:
\begin{equation}
f(x_P,t)=N~x_P^{1-2\alpha_P(t)}~ {B^2(t)\over 16 \pi} ~,
\label{flux}
\end{equation}
where  $B(t)$ describes the pomeron coupling to the proton and is parametrized
as below \cite{COLLINS}
\begin{equation}
B(t)= 4.6~mb^{1/2}~exp~(1.9GeV^{-2}~t) ~.
\label{bt}
\end{equation}
The normalization factor $N$ was set to be equal to
${2 \over \pi}$ following the convention of refs. \cite{DOLA1,DOLA2,COLLINS}.
In addition we have slightly increased
$\alpha_P(0)$ in the flux factor  $f(x_P,t)$ (\ref{flux})
up to $\alpha_P(0)=1.1$
 \cite{H12}.

The  functions $F^P_{2,L}(\beta, Q^2,t)$ are the pomeron structure
functions with the variable $\beta$ playing the role of the
Bjorken scaling variable for
the $\gamma^{*}(Q^2)$  pomeron inelastic "scattering".

 In the region
of large $Q^2$ the pomeron structure function $F_{2}^P$
is expected to be described in terms of
the QCD improved parton model and is related in the  conventional
way to the quark distributions $q_i^P(\beta,Q^2,t)$
in the pomeron:
\begin{equation}
F_{2}^P(\beta, Q^2,t) = 2~\beta~ \sum_i e_i^2~ q_i^P(\beta,Q^2,t) ~,
\label{pmod}
\end{equation}
where $e_i$ are the quark charges (note that $q_i^P= \bar q_i^P$. )
The above formula is given in the leading logarithmic approximation
of perturbative QCD in which the pomeron longitudinal
structure function  $F_L^P=0$.

At first we shall specify the details of the parton distributions in the
pomeron at the reference scale $Q_0^2=4GeV^2$.

At small $\beta$ both the quark and gluon distributions are assumed
to be dominated by the pomeron exchange .
\begin{eqnarray}
\beta~ q_i^P(\beta,Q_0^2,t)&=&a_i^P(t)~ \beta^{1-\alpha_P(0)} \nonumber \\
\beta~ g^P (\beta,Q_0^2,t)&=&a_g^P(t)~\beta^{1-\alpha_P(0)}
\label{smb}
\end{eqnarray}
The functions $a_i^P(t)$ and $a_g^P(t)$ can be estimated from the factorization
of
pomeron couplings \cite{DOLA1,BERGER,CAPELLA} :
\begin{eqnarray}
{}~~~~~a_i^P(t) = r(t)~a_i ~~~~~~~~~~~
a_g^P(t) = r(t)~a_g ~,
\label{factor2}
\end{eqnarray}
where the parameters $a_i$ and $a_g$
are the pomeron couplings
controlling the normalization of  the small $x$ behaviour of the
sea quark and gluon
distributions in the {\bf proton} i.e.
\begin{eqnarray}
xq_i(x,Q_0^2)+x\bar q_i(x,Q_0^2)&=&2~a_i~x^{1-\alpha_P(0)} \nonumber \\
xg(x,Q_0^2)&=&a_g~x^{1-\alpha_P(0)}
\label{smxp}
\end{eqnarray}
and the function $r(t)$ is:
\begin{equation}
r(t)={\pi\over 2}{G_{PPP}(t)\over B(0)}
\label{rt}
\end{equation}
The coupling $G_{PPP}(t)$ is the triple pomeron coupling  and
its magnitude can be estimated from the cross-section of the diffractive
production $p+\bar p \rightarrow p+X$ in the limit of large mass $M_X$
of the diffractively produced system $X$.
We neglected the (weak) $t$ dependence of the function $r(t)$ and have
estimated its
magintude from the recent Tevatron data \cite{TEVAT} as $r(t) \approx r(0)=
0.089$.  The parameters $a_i$ were estimated assuming that the
sea quark distributions in the {\bf proton} can be parametrized as:
\begin{equation}
 xq_i(x,Q_0^2)+x\bar q_i(x,Q_0^2)=2~a_i~x^{1-\alpha_P(0)}~(1-x)^7
\label{seap}
\end{equation}
and fixing the constants $a_i$ from the requirement that the average
momentum fraction which corresponds to those distributions is the same
as that which follows from the recent parametrization of parton distributions
in the proton \cite{MRSA,MRSG}.  The momentum sum rule has also been used to
fix
the parameter $a_g$ i.e. we assumed
\begin{equation}
xg(x,Q_0^2)=a_g~x^{1-\alpha_P(0)}~(1-x)^5
\label{glup}
\end{equation}
and imposed the condition that the gluons carry 1/2 momentum of the
 proton.
We extrapolated the pomeron dominated  quark and gluon  distributions in the
pomeron
(see (\ref{smb})) to the
region of
arbitrary values of $\beta$ by multiplying the factor $\beta^{1-\alpha_P(0)}$
 by $1-\beta$ \cite{BERGER}.

 We have also included the term
proportional to $\beta(1-\beta)$
in both the quark and gluon distributions \cite{BERGER}.  The normalization
of  this term in the quark distributions has been estimated in
\cite{DOLA2} assuming that it is dominated by the quark-box diagram
 with the non-perturbative couplings of pomeron to quarks.
In this model one gets:
\begin{equation}
\beta~ q^P(\beta,Q_0^2)={C \pi\over 3}~ \beta~(1-\beta) ~,
\label{land}
\end{equation}
where $C \approx 0.17$ \cite{DOLA2}.
We found that the fairly reasonable description of data can be achieved
provided that the constant $C$  is enhanced
by a factor equal to 1.5.  We have also assumed that the relative
normalization of the quark distributions in the  pomeron corresponding
to different flavours is the same as that of the sea quark distributions
in the proton \cite{MRSA,MRSG}.
Finally the normalization of the term proportional
to $\beta(1-\beta)$
in the gluon distribution in the pomeron has been obtained by imposing the
momentum sum rule.
Following the approximations discussed
above we have neglected the $t$ dependence in those parton distributions.

As the result of the estimates
and extrapolations discussed above the parametrization of parton
distributions in the pomeron at the reference scale $Q_0^2=$ 4GeV$^2$
looks as follows:
\begin{eqnarray}
\beta~ g^P(\beta,Q_0^2)&=&(0.218~ \beta^{-0.08} + 3.30 ~\beta)~ (1-\beta)
     \nonumber \\
\beta~ u^P(\beta,Q_0^2)&=&0.4~(1-\delta)~ S^P(\beta) \nonumber \\
\beta~ d^P(\beta,Q_0^2)&=&0.4~ (1-\delta)~ S^P(\beta) \nonumber \\
\beta~ s^P(\beta,Q_0^2)&=&0.2~ (1-\delta)~ S^P(\beta)  \nonumber \\
\beta~ c^P(\beta,Q_0^2)&=&\delta~ S^P(\beta) ~,
\label{ppar}
\end{eqnarray}
where the function $S^P(\beta)$ is parametrized as below
\begin{equation}
S^P(\beta)= (0.0528~\beta^{-0.08}~+~0.801~\beta)~(1-\beta)
\label{spar}
\end{equation}
and $\delta$=0.02 \cite{MRSA,MRSG}.
The  analysis of the pomeron structure functions based
on different parametrizations of parton distributions in the pomeron has
recently been presented in refs. \cite{CAPELLA,STIRLING}.

The parton distributions defined by eqs.(\ref{ppar},\ref{spar}) were next
evolved up to the values of $Q^2$
for which the data exist using the LO Altarelli-Parisi evolution equations
\cite{AP,APR} with $\Lambda=0.255$ GeV \cite{MRSG}.
In Fig.1 we show our results for the quantity:
\begin{equation}
 F_2^{D}(\beta,Q^2)=\int_{x_{PL}}^{x_{PH}}dx_P \int_{-\infty}^0 dt
{dF_{2}^D(x_P,\beta,Q^2,t)\over dx_P dt}
\label{dif0}
\end{equation}
with with $x_{PL}=0.0003$ and $x_{PH}=0.05$ \cite{H12},
plotted as the function of $x_P$ for different values of $\beta$
and $Q^2$.  We see a very good agreement with the recent data
from H1 collaboration at HERA  \cite{H12}. Dashed lines in Fig.1 show
$F_2^{D}(\beta,Q^2)$ computed when gluons were neglected
at the initial scale $Q_0^2=$ 4GeV$^2$.
%We see that gluons are indispensable at the initial scale
%to describe the data  at lower values of $\beta$.
%
\ffig{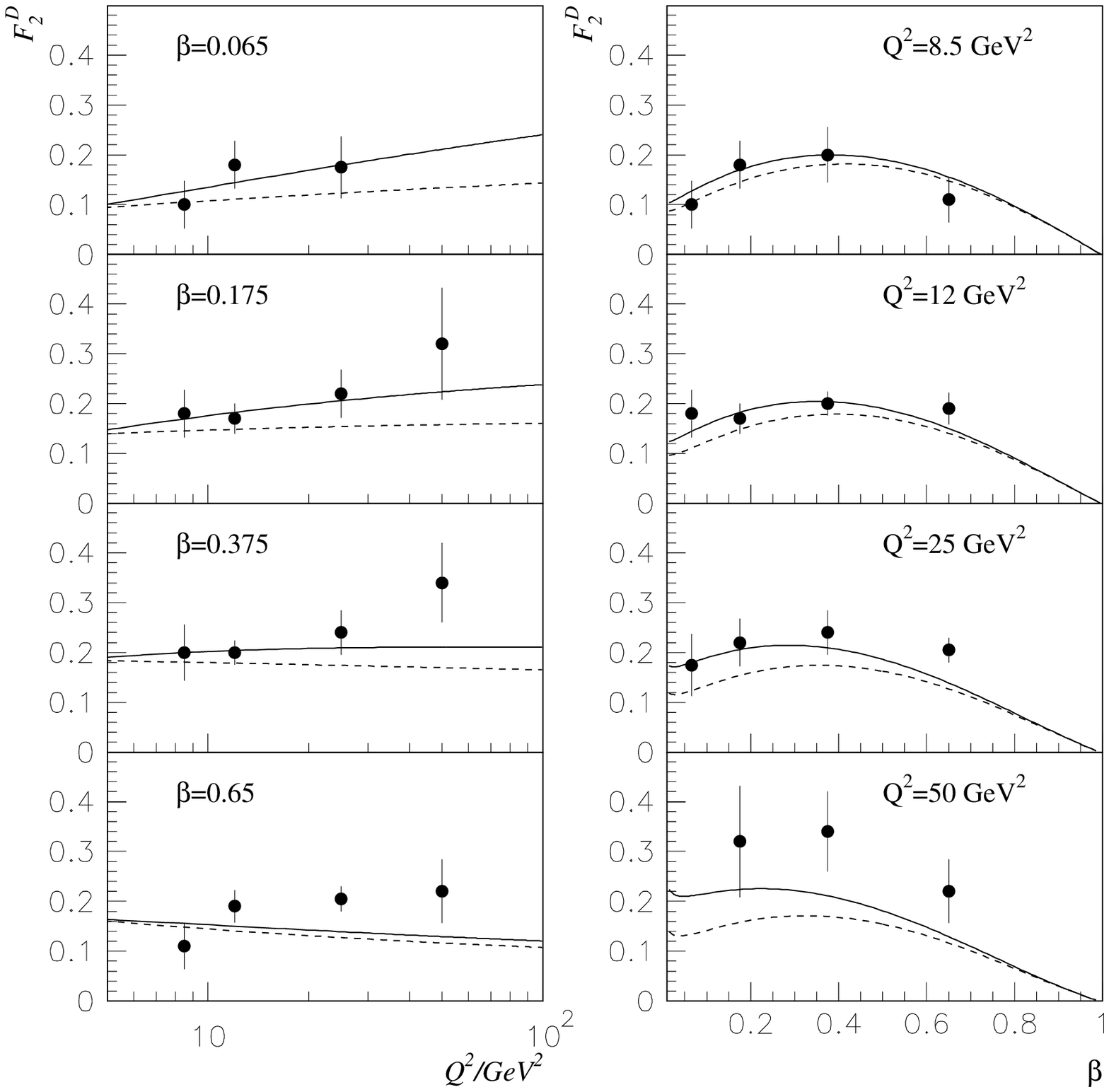}{90mm}{\em Theoretical predictions for diffractive
               $F_2^D(\beta,Q^2)$ and their comparison with the data
               from HERA (solid lines). Dashed lines show
               predictions when there are no gluons at the initial
               scale $Q_0^2$.}{fig1}

We have also performed the NLO QCD analysis in order to be able to consistently
introduce the quantity $R=F_L^D(\beta,Q^2)/F_T^D(\beta,Q^2)$ where
$F_T^D(\beta,Q^2)=F_2^D(\beta,Q^2)-F_L^D(\beta,Q^2)$ and
$F_L^D(\beta,Q^2)$ is defined in analogy to (\ref{dif0}).
We compared the NLO and LO results for $F_2^D(\beta,Q^2)$ from Fig.1 and found
that both approximations lead to similar predictions.
In Fig.2 we show our results for $R$.
\ffig{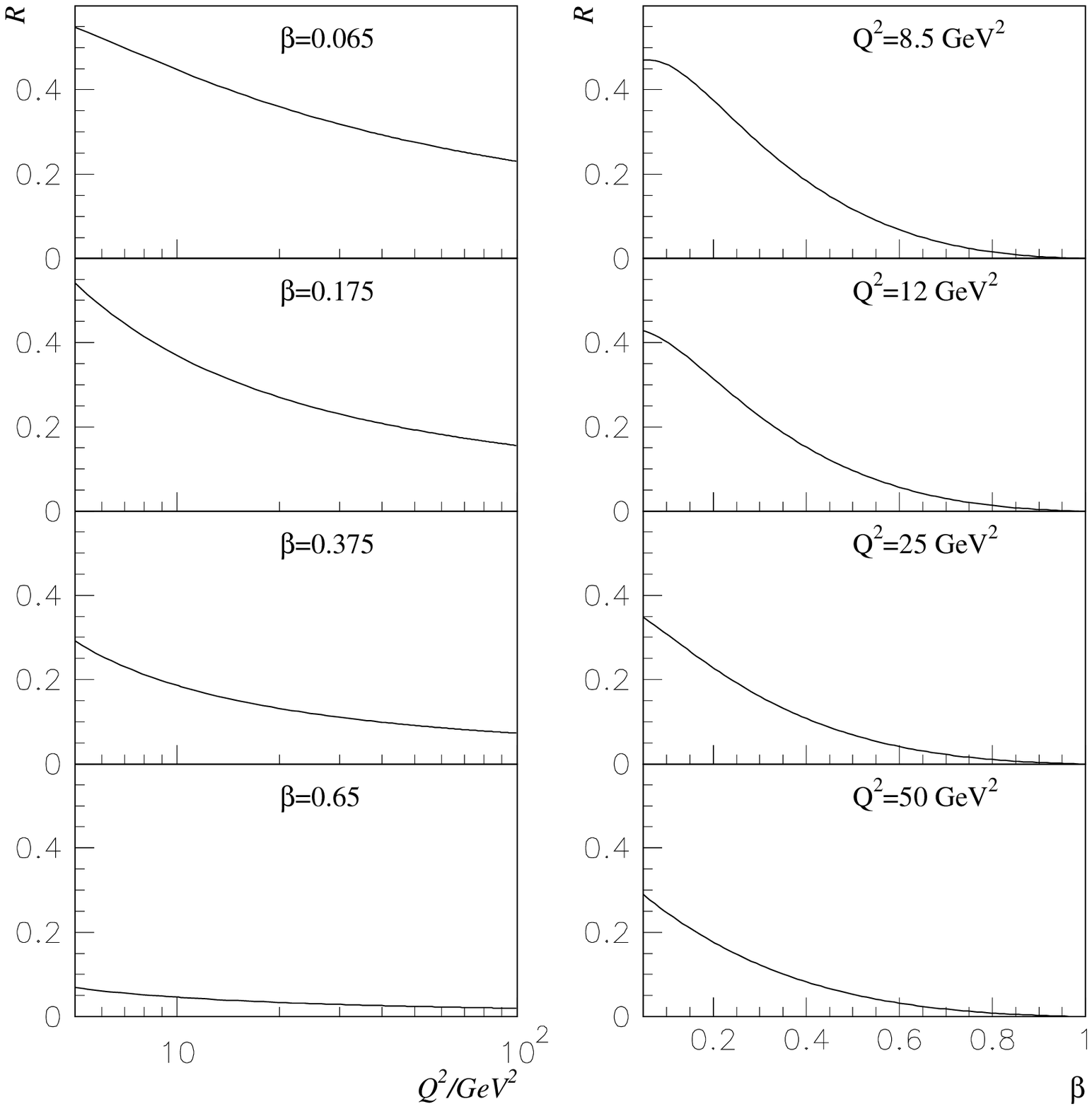}{90mm}{\em Theoretical predictions for diffractive R
      plotted as the function of $Q^2$ (for fixed $\beta$)
     and the function of $\beta$ (for fixed $Q^2$).}{fig2}
The longitudinal
diffractive structure function is driven mainly by the gluon distribution
in the pomeron and a large amount of gluons (see \ref{ppar}) implies that
$R$ can reach $0.5$ for $\beta \le 0.1$.
\\
\\
%
%\begin{center}
%{\large\bf Aknowledgements}
%\end{center}
We thank Albert de Roeck for several very instructive discussions which
prompted this study. This research has been supported in part by the
Polish State Committee for Scientific Research grants N0s 2 P302 062 04,
 2 P03B 231 08 and Maria Sk\l{}odowska-Curie Fund II (no. PAA/NSF - 94 - 158).

\end{document}